# *In situ* characterisation of nanoscale electromechanical properties of *quasi*-two-dimensional $MoS_2$ and $MoO_3$

Sumeet Walia[1,†] (✉), Hussein Nili[1,†], Sivacarendran Balendhran[1], Dattatray J. Late[2], Sharath Sriram[1], and Madhu Bhaskaran[1] (✉)

[1] Functional Materials and Microsystems Research Group, School of Electrical and Computer Engineering, RMIT University, Melbourne, Australia
[2] Physical and Materials Chemistry Division, CSIR-National Chemical Laboratory, Pune, India
† These authors contributed equally to this work

## ABSTRACT

Precise manipulation of electronic band structures of two-dimensional (2D) transition metal dichalcogenides and oxides (TMD&Os) *via* localised strain engineering is an exciting avenue for exploiting their unique characteristics for electronics, optoelectronics, and nanoelectromechanical systems (NEMS) applications. This work experimentally demonstrates that mechanically-induced electrical transitions can be engineered in quasi-2D molybdenum disulphide ($MoS_2$) and molybdenum trioxide ($MoO_3$) using an *in situ* electrical nanoindentation technique. It is shown that localised strains on such quasi-2D layers can induce carrier transport alterations, thereby changing their electrical conduction behaviour. Such strain effects offer a potential tool for precisely manipulating the electronic transport properties of 2D TMD&Os, and understanding the interactions of the atomic electronic states in such layered materials.

# 1 Introduction

The exotic electronic and optical properties of two-dimensional (2D) transition metal dichalcogenides and oxides (TMD&Os) are continuing to generate tremendous research interest [1-7]. The presence of an intrinsic bandgap and the possibility to manipulate their electronic energy states *via* a range of techniques underpins their potential and renders these materials highly versatile. Molybdenum disulphide ($MoS_2$) and molybdenum trioxide ($MoO_3$) are two of the most researched 2D TMD&Os, respectively [3]. In their bulk form, 2*H*-$MoS_2$ and *α*-$MoO_3$ are comprised of stacked planar crystals that are held together by weak Van der Waals forces. Such a unique structure allows their exfoliation into thin 2D layers of one to several unit cell thicknesses, analogous to graphene. These 2D layers exhibit starkly different electronic properties from their bulk counterparts and their electronic energy band structure are a function of the number of constituent layers [8,9]. These band structures can be tuned, and owing to their 2D morphologies, these crystals offer large surface areas that can be exploited for a plethora of electronic and optoelectronic devices such as transistors, catalysts, photovoltaic cells and nanosensors [3,9-17]. A much less attended paradigm in 2D materials relevant to thin films are quasi-2D materials made from the re-stacking of 2D flakes. In the re-stacking process of such 2D flakes, they can be placed on the top of each other and held together via Van der Waals forces, while retaining intrinsic electronic and optoelectronic properties of the constituent 2D layers.

A key feature of 2D $MoS_2$ and $MoO_3$ is that their electronic band structures are susceptible to manipulations using techniques such as substitutional doping, intercalation, solar irradiation and application of strains. Although, doping and intercalation methods have been widely employed for tuning the electronic energy states of TMD&Os, a precise control over such alterations is yet to be demonstrated [3,7]. As a less investigated electronic band structure manipulation method, the application of localized strain is a potentially attractive approach for tuning the band gaps and electronic transport properties of TMD&Os monolayers. It has even been shown that by applying such localized strains, metal–semiconductor transitions can be achieved [18]. However, such strain effects are not fully understood, and a critical understanding of lattice structure mediated carrier transport, is needed to harness them. The strain effect can also potentially affect the properties of the re-stacked flakes. As a result, the thin film, as a whole, can show electronic and mechanical properties similar to that of the 2D flakes. There are several theoretical studies demonstrating that the electronic structure of 2D $MoS_2$ and $MoO_3$ can be engineered by applying strains due to the compressions within the monolayer inducing geometric changes [19-23]. However, there are no experimental electromechanical results reported to confirm these predictions. Some recent atomic force microscopy (AFM) indentation experiments have shown that monolayer $MoS_2$ can withstand high magnitude of strain without exhibiting bond breakage [24]. However, the conductance of the monolayers is not directly acquired in such experiments due to the effects of contact resistance as well as the influence of the metallic contacts on the band structures of the monolayers. Therefore, it is imperative to consider these effects and investigate the electron conduction through the 2D structures to investigate their viability for nanoelectromechanical systems (NEMS).

In this work, we analyse the effect of strain on the electronic properties of 2D $MoS_2$ and $MoO_3$ as model TMD&Os, by using a nanoindentation technique with unique *in situ* electrical characterisation. This technique allows the study of charge transport mechanisms and functional properties of complex material systems under highly localised strains at the nanoscale [25-27]. and offers an exceptional capability of real time observations of indentation induced electromechanical phenomena. The system allows for a precise control of the tip's penetration depths, providing a highly accurate load and depth measurement. The technique can be imagined as a tip being used to strain the surface of an inflated balloon. Up to a certain strain level, the balloon retracts to its original position on the removal of the applied strain (completely elastic). However, it shows surface damage when strain exceeds a certain threshold (plastic deformation), and bursts beyond a point (fracture). Comprehensive details about the technique are provided in several articles that exist in literature [25,26,28-30].

# 2 Experimental

Our work aims to show an effect of strain on quasi-2D thin films that are composed of re-stacked 2D nanoflakes of $MoS_2$ and $MoO_3$, respectively. The *in situ* nanoindentation tests are performed on 200-300 nm thick $MoS_2$ and $MoO_3$ quasi-2D films that comprise of few layered (2-7 layers of $MoS_2$ and majority 2 layers of $MoO_3$) liquid-exfoliated 2D nanoflakes of the two respective materials deposited onto substrates with electrically conductive fluorine-doped tin oxide (FTO) coating (see Methods section for details).

Address correspondence to Sumeet Walia, waliasumeet@gmail.com; Madhu Bhaskaran, madhu.bhaskaran@gmail.com

# 3 Results and discussion

High resolution transmission electron microscopy (HRTEM) was conducted in order to verify the crystal structures of the as-synthesised 2D $MoS_2$ and $MoO_3$ nanoflakes. HRTEM images for $MoS_2$ and $MoO_3$ nanoflakes are shown in Figs. 1(a) and 1(b), respectively. The lattice fringes for $MoS_2$ of 0.27 nm and $MoO_3$ of 0.39 nm (insets of Figs. 1(a) and 1(b), respectively) are consistent with the presence of planar $2H$-$MoS_2$ and $\alpha$-$MoO_3$ [1,6,10,16]. An AFM image of a sample cross-section of a $MoS_2$ thin film (Fig. 1(c)) show that the nanoflakes are oriented parallel to each other (see section S1 in ESM for a corresponding 3D image). This alignment can be ascribed to the well-known Van der Waal interactions, liquid surface tension and the centre of gravity that facilitate the nanoflakes to arrange themselves in parallel [31-33].

Such an arrangement is also aided by the relatively small lateral dimensions of the nanoflakes as evidenced by dynamic light scattering experiments (DLS) reported for both 2D $MoS_2$ [34] and $MoO_3$ [1] nanoflakes in separate studies. Such a well-stratified arrangement of layers ensures that the nanosheets are stacked uniformly to form multilayers that preserve their 2D characteristics.

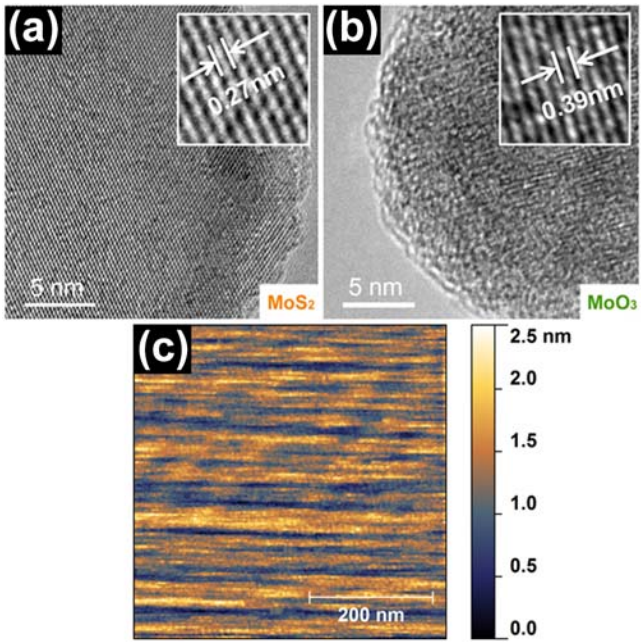

**Figure 1** Transmission electron micrograph of a liquid exfoliated 2D nanoflakes of (a) $MoS_2$ and (b) $MoO_3$ with the inset corresponding to its high resolution transmission electron micrographs. (c) Atomic force microscopy cross-sectional scan of a $MoS_2$ thin film composed of 2D nanoflakes. It shows that the nanoflakes are aligned parallel to each other.

We refer the readers to our previous publications for more extensive characterizations of the 2D $MoS_2$ and $MoO_3$ nanoflakes used in this study [1,6,10,16].

For the proposed strain engineering in 2D layers, inelastic strain relaxation as a result of dislocation plasticity or fracture must be avoided. Therefore, prior to investigating the effects of strain on the electronic conduction properties of the 2D $MoS_2$ and $MoO_3$ layers, their nanomechanical properties were determined. This data ensured that the applied force results in minimal plastic deformation.

The nanomechanical properties are determined using a conductive Berkovich indenter tip [26,35]. The tip is used for applying load cycles at multiple locations on the quasi-2D films, while their load–displacement characteristics are recorded. The Young's modulus and the hardness are extracted from ten indents at different points on the surface. The elastic modulus and hardness of the films made of 2D $MoS_2$ nanoflakes are on average 193.5±4.6 GPa and 11.5±0.4 GPa, respectively, whereas the corresponding values are 176.3±4.1 GPa and 10.9±0.4 GPa for films comprising of 2D $MoO_3$ nanoflakes. Previously reported values of Young's modulus for 2D $MoS_2$ range between 130-270 GPa [36,37], while the nanomechanical properties of 2D $MoO_3$ have yet to be reported to our knowledge. Variations in comparison to previously reported values can be expected due to the difference in the synthesis procedures and the quasi-2D nature of these materials.

To experimentally determine the effect of strain on the electrical characteristics, we strained the as-synthesised 2D layers of $MoS_2$ and $MoO_3$ using a Berkovich indenter tip (see Experimental section for details) and acquired *in situ* current–voltage (*I–V*) characteristics. The measurements were performed for loads of 50 μN to 6 mN. The loads are selected such that there is no substantial plastic deformation and a true effect of strain can be observed. The ability to perform *in situ* electrical nanoindentation for varying nanocontact sizes is a significant capability that allows to probe a given nanocontact area at varying loads and reveal localised *in situ* nanoelectromechanical behaviour.

Constant bias *in situ* electrical nanoindentation experiments were performed on multiple arrays of more than hundred indents on different areas across various samples. It is ensured that the indentation process does not compromise the structural integrity of the nanoflakes.

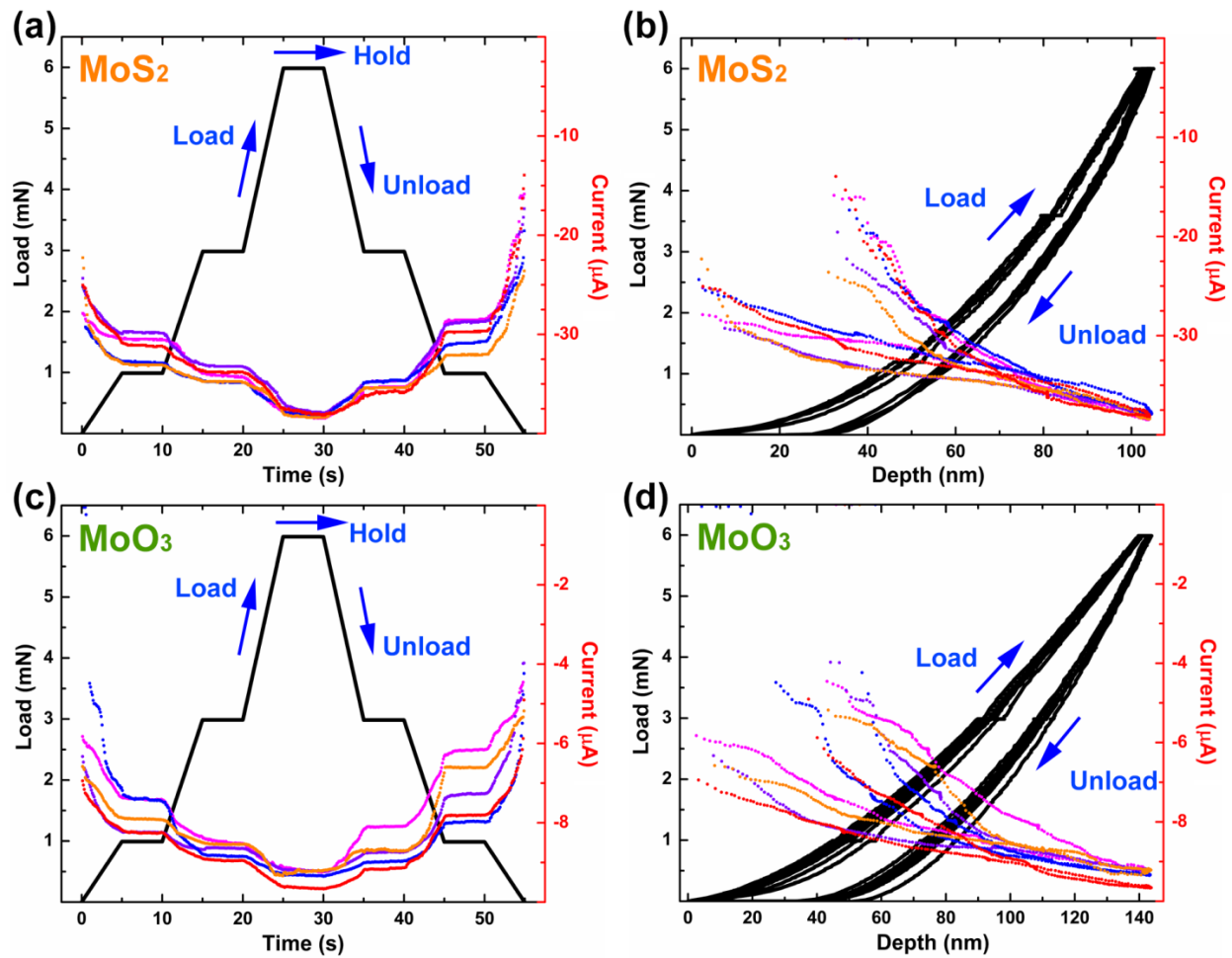

**Figure 2** (a, b) Time-based correlation between the contact load and current at constant bias, correlation between the nanomechanical and the electrical characteristics of nanocontacts respectively for the films made of 2D $MoS_2$ and (c, d) for 2D $MoO_3$ nanoflakes. The multiple curves represent the characteristics acquired at different points on the samples to show the consistency of the strain effects.

The dynamical evolution of two or more inter-dependent dynamic responses should be correlated by an independent (or neutral) variable (e.g. time, frequency) in any measurement setup. The evolution of inter-dependent variables as a function of the independent variable provides significant information about the consistency of the response and is the basis of cross-correlation analysis between two or more mutually impacting phenomena. As such, time-based current correlation for five representative points on the films made of $MoS_2$ and $MoO_3$ nanoflakes is shown in Figs. 2(a) and 2(c), respectively. A constant bias of –0.1 V was used for these measurements. As can be seen, the load was gradually increased and sequentially held at 1, 3, and 6 mN respectively, to estimate the nanocontact area at varying load levels. A larger current is observed for the $MoS_2$ sample due to its lower intrinsic bandgap. It can be seen that the magnitude of current progressively increases with load and follows a consistent trend for both $MoS_2$ and $MoO_3$ films at different points across the samples.

The current remains constant at zero load gradients, confirming that the sample does not undergo any significant thermodynamical variations under large mechanical pressures. Figs. 2(b) and 2(d) show the actual nanocontact current correlated with the nanomechanical behaviour of the corresponding indented areas. These current vs. depth/load curves (Figs. 2(b) and 2(d)) clearly reveal that the current drops back to zero at full-unload (see Section S1 in the Electronic Supplementary Material (ESM) for a detailed description of the term "full-unload"), eliminating the possibility of any substantial plastic deformation of the 2D layers. Hence, these results provide a comprehensive depth/load-based correlation of nanoelectromechanical behavior, by demonstrating the real-time current change at varying strain levels. Additionally, displacement controlled partial load/unload experiments were also performed to further study the nanoscale electromechanical characteristics.

The contact diameter and the evolution of the nanocontact area at varying strain magnitudes for both

2D MoS$_2$ and MoO$_3$ are estimated (see Section S3 in the ESM). Based on the aforementioned estimates, target contact depth points for both materials are chosen such that similar force magnitudes are required to attain these depths on MoS$_2$ and MoO$_3$. Figures 3(a,b) and 4(a,b) shows the partial load/unload patterns and the corresponding depths obtained for indents on MoS$_2$ and MoO$_3$ nanoflakes, respectively. The corresponding load/displacement curves are shown in Figs. 3(c) and 4(c), respectively. It can be seen that, at a given depth, the load is always higher during the loading segment compared to the corresponding depth in the unloading regime. *In situ* I–V characteristics were obtained by applying range bound voltage sweeps (–0.5 to 0.5 V) at varying constant force magnitudes (50 μN to 6 mN) to reveal the effect of strain on the electrical conduction in the layers made of 2D nanoflakes. The nanoindentation system used for the measurements is set up such that the indenter tip grounded, with bias voltage applied to the bottom FTO electrode. I–V characteristics of bare FTO show it is perfectly ohmic under different loads and eliminates the possibility of any substrate effects on the electrical measurements (see Section S4 in ESM). The load–depth and load–time curves shown in Figs. 2, 3 and 4 clearly reveal a high degree of elasticity (indicated by the return of the unloading curve close to the starting position) and repeatability in the electromechanical response obtained on multiple arrays of indents over a large area.

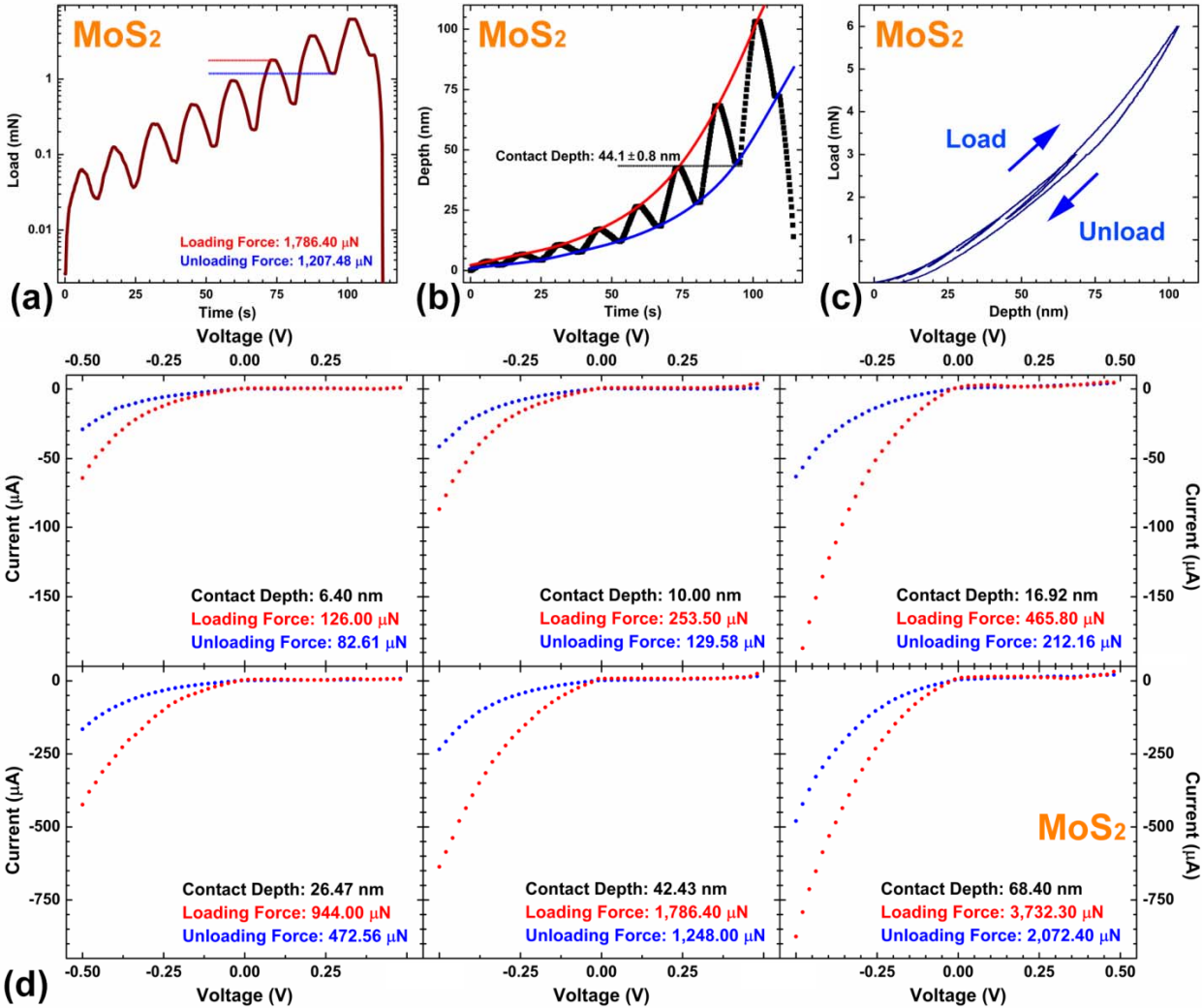

**Figure 3** Representative curves for displacement controlled partial load/unload experiments. (a) Loading profile, (b) depth profile, (c) corresponding load–displacement curve for *in situ* nanoindentation on 2D MoS$_2$ nanoflakes, and (d) I–V characteristics under loading and unloading at varying contact depths.

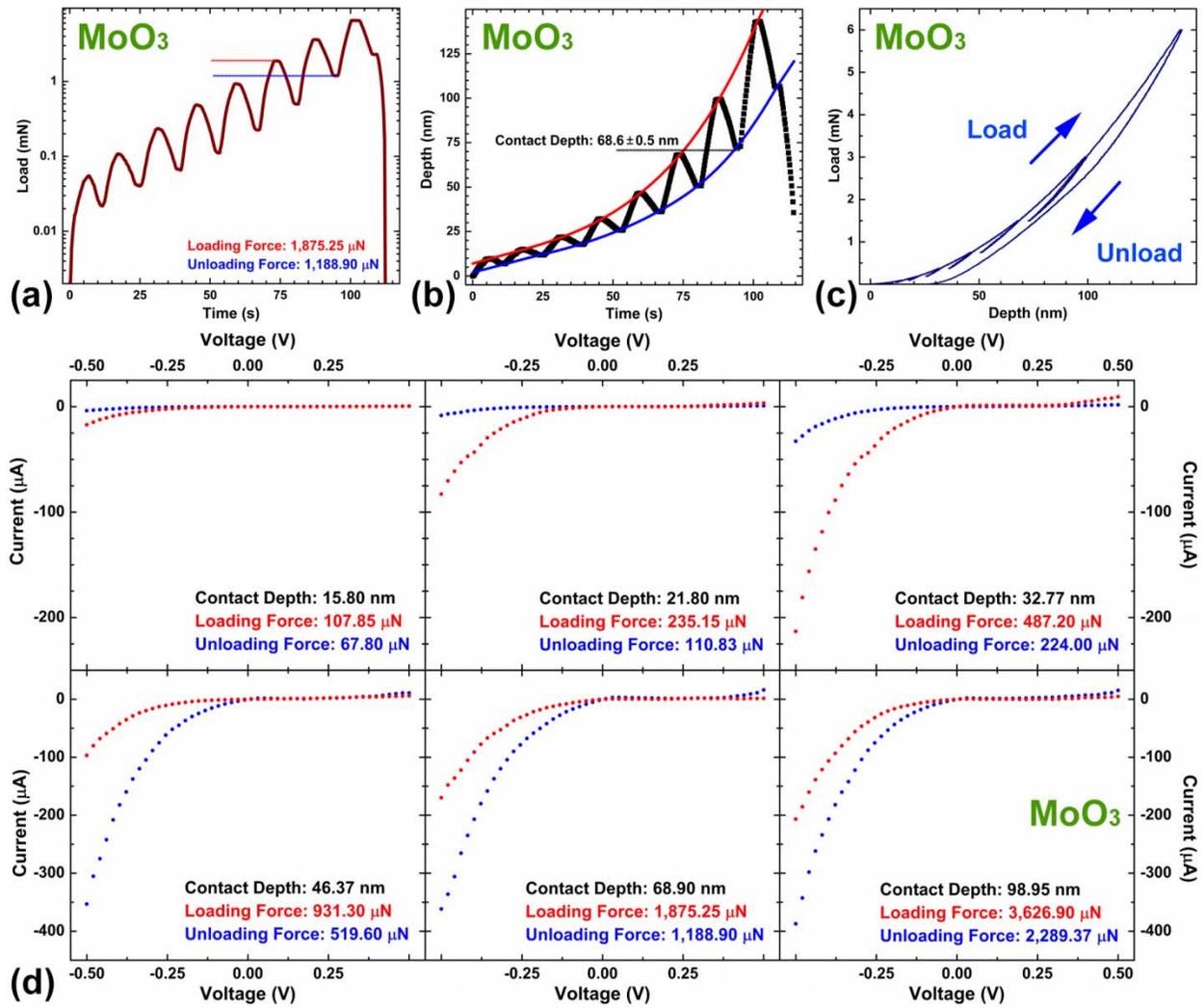

**Figure 4** Representative curves for displacement controlled partial load/unload experiments. (a) Loading profile, (b) depth profile, (c) corresponding load–displacement curve for *in situ* nanoindentation on 2D MoO₃ nanoflakes, and (d) I–V characteristics under loading and unloading at varying contact depths.

Highly repeatable nanoelectromechanical characteristics (Figs. 2(b) and 2(d)) of multiple indents prove that deformations occurring at the strain levels used in this study are largely elastic and reversible (see Section S3 in ESM for representative load–displacement curves for typical elastic and plastic deformation). These experimental characteristics along with the parallel alignment of the nanoflakes provide fundamental evidence that there is minimal nanoflake movement or layer slipping and that the mechanical alterations in the nanoflakes during the indentation process are largely reversible, which may otherwise create data distortions. Based on this, it can also be hypothesised that the gap between the nanoflakes does not play a crucial role.

Now we discuss the role played by the Schottky barrier at the tip/nanoflake interface. The conductive Berkovich indenter tip comprises of a vanadium carbide coating. The work function of vanadium carbide is estimated to be 5.02 eV [38], while 2D MoS$_2$ exhibits a work function of approximately 5.20 eV [39]. This combined with the fact that a perfectly ohmic contact is observed between the indenter tip and the FTO substrate (work function of 4.4 eV)[40], under varying loads (see Section S4 in the ESM) eliminates the possibility of the Schottky barrier having any significant impact on the transport characteristics in this case.

The I–V characteristics of both 2D MoS$_2$ and MoO$_3$ (Figs. 3(d) and 4(d), respectively) at extremely shallow to high depths show that an increasing load results in a higher current. In these experiments, the role of the contact resistance has been eliminated by determining the change in current due to increase in contact area and normalising all I–V curves to eliminate this contribution. As such, the *in situ* electrical characteristics reveal the actual effect of applied strain on the electrical characteristics of the 2D layers. Multiple I–V sweeps at the same location after releasing and subsequently reapplying strain resulted in similar characteristics, further confirming that the 2D layers experience minimal plastic deformation. Expectedly, higher loads at identical contact depths result in larger currents. Even at extremely low contact depths (low load), it is observed that the current shows striking dependence on load (see Tables 1 and 2). At higher loads, when the resistivity at the nanocontact interface is significantly lower, a large change in current is still observed under different loads at identical contact diameters (see Tables 1 and 2). This proves that the alteration of the current profile is the result of change in the material's electronic properties and not the effect of the nanocontact interfacial characteristics. Also the fact that the electrical properties of the nanocontact are fully characterized on standard conductive samples, and displays an ohmic behaviour assures that the contact electrical properties can be isolated from the inherent response of the 2D layers.

The non-linearity of the stress-field in the indentation experiments ensures that electronic structure alterations are non-uniform. Although, the contact plane and the underlying substrate complicate the strain states that are achieved in our experiments, the elastic nanomechanical response strongly suggests that these are not the dominant factors. The anisotropic nature of the I–V characteristics can be attributed to this non-uniform nature of the indentation testing which also causes the atomic level strain to be non-uniform.

**Table 1:** Summary of *in situ* electrical nanoindentation results for thin films made of 2D MoS$_2$ nanoflakes.

| Contact depth (nm) | Loading force (μN) | Unloading force (μN) | Load/ Unload ratio | % difference in current |
|---|---|---|---|---|
| 6.40 ± 0.3 | 126.00 | 82.61 | 1.52 | 120.61 |
| 10.00 ± 0.5 | 253.50 | 129.58 | 1.95 | 111.35 |
| 16.92 ± 0.4 | 460.85 | 212.16 | 2.17 | 227.36 |
| 26.47 ± 0.5 | 944.00 | 472.56 | 1.99 | 156.21 |
| 42.43 ± 0.8 | 1786.40 | 1248.00 | 1.43 | 171.78 |
| 68.48 ± 1.0 | 3732.30 | 2072.40 | 1.80 | 82.75 |

**Table 2:** Summary of *in situ* electrical nanoindentation results for thin films made of 2D MoO$_3$ nanoflakes.

| Contact depth (nm) | Loading force (μN) | Unloading force (μN) | Load/ Unload ratio | % difference in current |
|---|---|---|---|---|
| 15.80±0.5 | 107.85 | 67.80 | 1.59 | 426.38 |
| 21.80±1.0 | 235.15 | 110.83 | 2.12 | 891.86 |
| 32.77±1.0 | 487.20 | 224.00 | 2.17 | 549.08 |
| 46.37±2.0 | 931.30 | 519.60 | 1.79 | 264.74 |
| 68.80±1.5 | 1875.25 | 1188.90 | 1.57 | 112.94 |
| 98.95±2.0 | 3626.90 | 2289.37 | 1.58 | 87.40 |

The observed reduction in resistance (or enhanced current) with increasing load indicates that the mechanical strain causes alterations in the electronic structure of the 2D materials. Theoretical and computational studies of strain effects on 2D MoS$_2$ and MoO$_3$ have been widely reported in literature and have determined that strain modifies electronic states, potentially altering their bandgaps [20-23]. The highly elastic response as evidenced by Figs. 3c and 4c indicate that in our experiments this alteration is largely reversible even after several indents at the same spot. Simulations of the partial DOS suggest that the molybdenum centres are the major contributors to both valence and conduction bands in 2D MoS2. All significant changes in the band structure of 2D MoS2 on the application of strain can be related to shifts in the energy states of the valence band and the conduction band.

The strain changes the bond lengths as well as distances between individual atoms [23,24]. As a result, the application of strain shifts the energy states. Extensive simulations (reported elsewhere) of the partial density of states reveal that these states originate largely from the 3p orbitals of S atoms and 4d orbitals of Mo atoms. The applied strain alters the distances between the atoms which results in an atomic orbital overlap, leading to shifts in the energy of these states (dominated by the states of the valence band near the $\Gamma$ point and the conduction band between the $K$ and $\Gamma$ points) [14,23,24]. These observations are further supported by an atomistic tight-binding model that was developed by Gomez *et al.* to predict the effect of inhomogeneous strain on the local electronic states in 2D $MoS_2$ [41]. The phenomenon of strain induced electronic transitions has also been recently demonstrated in single crystalline $MoS_2$ and shows a reduction in bandgap with strain [42]. The consequent changes in the electronic band structures are expected to change the conducting nature of 2D $MoS_2$ layers [14,23,24,42-46]. Similarly, DOS changes due to applied strain are also expected to alter the electronic band structure of 2D $MoO_3$. A recent study on $MoO_3$ monolayers predicts that application of c-axis strain is expected to shrink the bandgap [20]. It is obvious that the magnitude of applied strain can be used to progressively manipulate the energy gap, which changes the conduction nature of the system, as evident from the $I$–$V$ characteristics demonstrated in this work. These observations are in-line with theoretical predictions and simulations that are extensively reported in literature and represent the first experimental validation of those predictions. These comprehensive observations demonstrate that the electrical transport characteristics are indeed modulated by localised strain. Furthermore, it is also seen that under similar force magnitudes, the contact depth attained on $MoS_2$ nanoflakes is lower, corroborating our earlier observation that $MoS_2$ is mechanically harder than $MoO_3$ nanoflakes. As such, strain appears to be a viable tool for creating dynamically varying bandgap profiles in 2D TMD&O membranes.

## 4 Conclusions

In summary, we have shown that the application of strain can change the electronic conduction characteristics of 2D TMD&Os. The unique re-stacking of the 2D nanoflakes to create quasi-2D films can potentially provide a new method for synthesising Van der Waal heterostructures. A state-of-the-art nanoindentation system was used to perform *in situ* electrical tests after eliminating the role of the contact resistance. It is demonstrated that mechanical strain offers a potential tool for controllably tuning the transport properties of 2D TMD&Os, thereby adding to their operational versatility. Such electromechanical tunability is highly desirable due to potential applications in novel NEMS, for instance NEM switches have been projected as a substitute technology for potential ultralow-power digital integrated circuits on the basis of zero off-state current and abrupt on/off switching advantage, which can offer zero standby power utilization and enormously steep switching (<0.1 mV/decade) [47]. Furthermore, the possibility of generating large strain-induced variations in electrical characteristics opens the door for a variety of applications in atomically thin systems including photovoltaics, quantum optics, nanosensors, and two-dimensional optoelectronic devices.

## 5 Experimental

### 5.1 Synthesis of quasi 2D $MoS_2$ and $MoO_3$ nanoflakes

A liquid exfoliation technique was employed to synthesize the two-dimensional (2D) $MoS_2$ and $MoO_3$ nanoflake suspensions [1,6,16]. 0.5 mL of N-methyl-2-pyrrolidone (NMP) (99% anhydrous, Sigma Aldrich) was added to 1 g of commercial $MoS_2$ and $MoO_3$ powder (99% purity, Sigma Aldrich) separately and the mixture was subsequently mechanically ground in a mortar for 30 min. The mixture was allowed to dry overnight under ambient conditions. The as-dried mixture was re-dispersed in 40 mL ethanol/water (1:1) solution. This solution was probe-sonicated (Ultrasonic Processor GEX500) for 90 min at a power of 125 W and the supernatant containing the exfoliated quasi 2D $MoS_2$ and $MoO_3$ nanoflakes was separated after being centrifuged for 45 min at the speed of 4000 rpm. Comprehensive characterizations of the as liquid exfoliated quasi 2D $MoS_2$ and $MoO_3$ nanoflakes have been reported in our previous works [1,6,16,34].

## 5.2 Transmission electron microscopy

TEM and HRTEM images of the $MoS_2$ and $MoO_3$ nanoflakes were acquired using a JEOL 2100F HRTEM.

## 5.3 Nanoindentation with *in situ* electrical characterisation

A Hysitron Triboindenter, configured to perform *in situ* electrical measurements using a conductive Berkovich tip was used in this work. The Berkovich tip is a three-sided pyramid with a total included angle of 142.3º and a half angle of 65.35º, and is considered the standard for nanoindentation measurements. *In situ* electrical measurements were performed using the NanoECR system [25,26]. Testing was performed at room temperature and pressure. The tip was grounded in NanoECR mode for all voltage and current measurements.

## Acknowledgements

The authors acknowledge support from the Australian Research Council (ARC) with project support through Discovery Projects DP130100062 (S.S.) and DP140100170 (K.K.-Z and M.B.); Australian Post-Doctoral Fellowships through Discovery Projects DP1092717 (M.B.) and DP110100262 (S.S.); and equipment funding through the Linkage, Infrastructure, Equipment, and Facilities Grant LE110100223. The authors thank Yichao Wang and Matthew Field for assistance with transmission electron microscopy.

**Electronic Supplementary Material**: Supplementary material (3D cross sectional AFM image, contact area, representative load–displacement curves to show difference between plastic and elastic deformation and *I*–*V* characteristics of the bare FTO substrates) is available in the online version of this article at http://dx.doi.org/10.1007/s12274-***-****-* (automatically inserted by the publisher).

# Electronic Supplementary Material

# *In situ* characterisation of nanoscale electromechanical properties of *quasi*-two-dimensional MoS$_2$

Sumeet Walia[1,†] (✉), Hussein Nili[1,†], Sivacarendran Balendhran[1], Dattatray J. Late[2], Kourosh Kalantar-zadeh[1], Sharath Sriram[1], and Madhu Bhaskaran[1] (✉)

[1] *Functional Materials and Microsystems Research Group, School of Electrical and Computer Engineering, RMIT University, Melbourne, Australia*
[2] *Physical and Materials Chemistry Division, CSIR-National Chemical Laboratory, Pune, India*
[†] *These authors contributed equally to this work*

## S1. 3D Image of a Cross-Section of Thin Films Composed of MoS$_2$ Nanoflakes

Figure S1 shows that the nanoflakes are oriented in a parallel re-stacked configuration as described in the main manuscript.

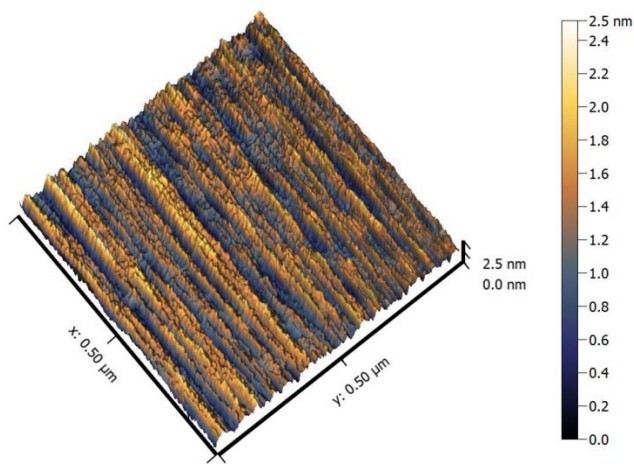

**Figure S1** Three-dimensional (3D) view of a thin film cross-section composed of MoS$_2$ nanoflakes.

## S2. Characterization of Nanocontact Properties for Accurate Determination of the Electromechanical Response of 2D MoS₂ and MoO₃ Nanoflakes

In a typical nanoindentation experiment, the indenter probe is held constant with a pre-load (usually a few µN) for a defined period of time to account for the surface friction effects and minimize thermal drift. During this period, the probe forms a pseudo-elastic contact with the sample, the diameter of which can be accurately calculated based on the geometry of the probe and basic Hertzian contact equations. All qualitative/quantitative assessment of the indentation data is carried out while the indenter probe is in contact with the sample. The "full-unload" term corresponds to a lower threshold in the unloading response curve (from ~15-20% of the maximum load for hard and brittle compounds to ~2-5% for flexible samples), where the indentation contact area can be assumed as the final contact area for a certain mechanical load after the elastic component of mechanical response has ceased. A fully stable and verifiable ultimate contact area is existent at that point. To ensure an accurate assessment of in situ electrical nanoindentation results, the properties of the nano-contacts formed by the conductive indenter tip have to be precisely chartered. The evolution of the contact depth is a direct function of the mechanical properties of the indented materials. We have characterized the electrical and mechanical properties of the nano-contacts through nanoindentation experiments on standard samples. A standard fused quartz sample was used to characterize the projected nano-contacts area subject to contact depth while standard bulk gold and platinum thin films were used to evaluate the nano-contacts resistance through in situ electrical measurements.

The projected contact area of the conductive Pulsar™ Berkovich tip (with an estimated elastic modulus of 470~500 GPa, Poisson's Ratio of 0.22 and an average resistivity of 247.6 µΩ.cm) used in this study [S1], can be described as a function of contact depth (hc):

$$A_c = C_0 h_c^2 + C_1 h_c + C_2 h_c^{1/2} + C_3 h_c^{1/4} + C_4 h_c^{1/8} + C_5 h_c^{1/16} \tag{S1}$$

where $C_n$ represents the fitting constants calculated based on the measured contact areas on the standard fused quartz sample. Figure S2 shows the calculated projected area function of the Berkovich tip along with the estimated nano-contact areas at constant contact depths on MoS₂ and MoO₃ nanoflakes.

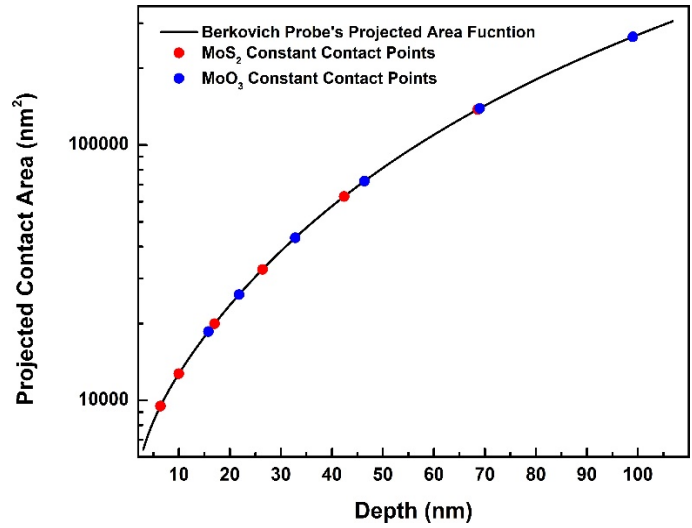

**Figure S2** The projected contact area of the Berkovich indenter tip at different contact depths on the 2D MoS$_2$ and MoO$_3$ nanoflakes.

Moreover, the nominal contact diameter (d) at a given penetration depth ($h$) for an elasto-plastic nano-contact can be explained in terms of a geometric contact [S2,S3]:

$$d = 2 \times \sqrt{2R_e h - h^2} \tag{S2}$$

where the effective radius of the probe (Re) is calculated as the average between the radii of indenter's rounding sphere (186±4 nm) and the radius of the circle circumventing the projected contact area. Figure S3 illustrates the evolution of nano-contact diameters versus nanoindentation loads in displacement controlled loading and unloading segments for MoS$_2$ and MoO$_3$ nanoflakes.

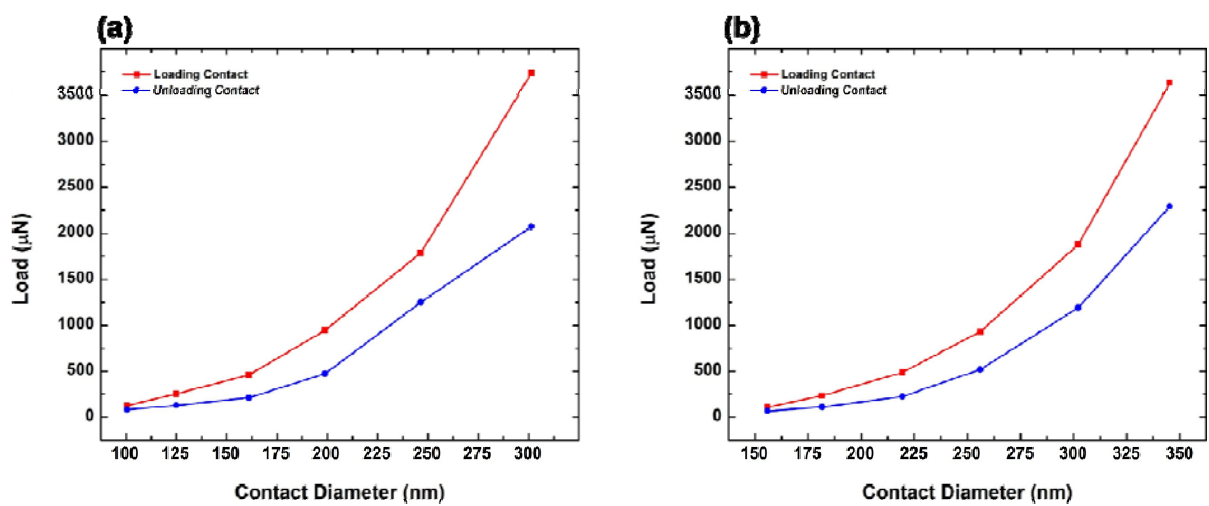

**Figure S3** The loading and unloading contact diameters for 2D nanoflakes of **(a)** MoS$_2$ and **(b)** MoO$_3$.

## S3. Typical Load–Displacement Curves for an Elastic and Plastic Deformation

Figure S4 represents the difference between load-displacement curves of elastic and plastic deformation. The average shape of the curves obtained in our study is highly reproducible and resemble the typical curves that are obtained in case of elastic deformation. The significance of elastic loading and unloading is that an apparent force equilibrium is set up between the sample and the indenter which makes the indenter behave as if it were in an elastic medium and this behaviour is indicated on the load-displacement curve.

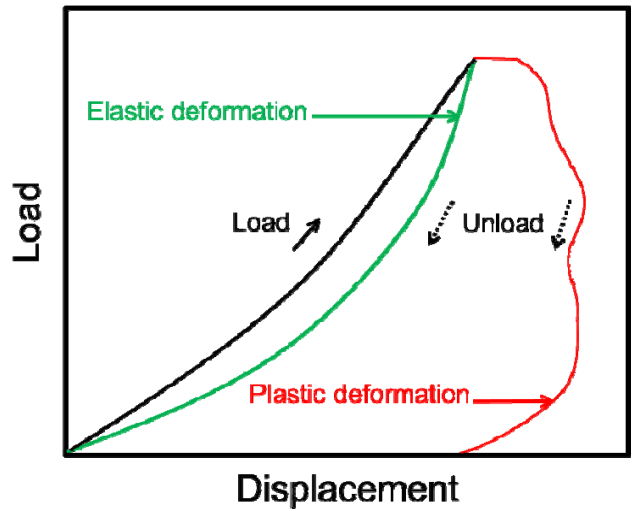

**Figure S4** Representation of typical load-displacement curves indicating elastic and plastic deformation.

## S4.  *I–V* Characteristics of the FTO Substrate

Indents on bare FTO substrate were performed under different force magnitudes. This is done to reveal the *I–V* behaviour of the substrate under strain and to show that the *I–V* characteristics of 2D $MoS_2$ and $MoO_3$ nanoflakes are unaffected by substrate effects. Figure S5 shows that the FTO substrate is perfectly ohmic in nature.

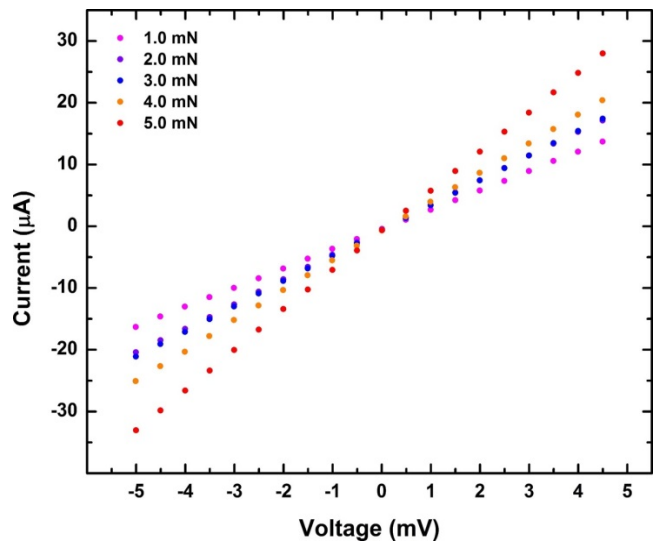

**Figure S5** *I–V* characteristics of the FTO substrate under varying force magnitudes demonstrating the perfectly ohmic nature of the Berkovich indenter's tip nano-contacts.

Address correspondence to Sumeet Walia, waliasumeet@gmail.com; Madhu Bhaskaran, madhu.bhaskaran@gmail.com